\documentclass[aps,amssymb,amsmath,twocolumn, pra]{revtex4}
\usepackage{graphicx}
\usepackage{epsfig}
\usepackage{dcolumn}
\usepackage{bm}
\usepackage{bbm}
\usepackage[up]{subfigure}
\newcommand{\be}{\begin{equation}}
\newcommand{\ee}{\end{equation}}
\newcommand{\bc}{\begin{center}}
\newcommand{\ec}{\end{center}}
\newcommand{\bea}{\begin{eqnarray}}
\newcommand{\eea}{\end{eqnarray}}

\begin{document}
\title{Implementing the one-dimensional quantum (Hadamard) walk using a\\
 Bose-Einstein condensate}
\author{C. M. \surname{Chandrashekar}}%
\affiliation{Atomic and Laser Physics,
             University of Oxford,
             Oxford OX1 3PU, United Kingdom}

\begin{abstract}


We propose a scheme to implement the simplest and best-studied version
of the quantum random  walk,  the  discrete Hadamard  walk,  in one dimension
using a coherent macroscopic  sample of ultracold atoms, Bose-Einstein
condensate  (BEC).  Implementation  of the quantum walk  using a BEC  gives
access to the familiar quantum  phenomena on a macroscopic scale. This
paper  uses a rf pulse  to implement the Hadamard operation  (rotation) and
stimulated Raman  transition technique as a unitary  shift operator. The
scheme suggests the implementation of the Hadamard operation and unitary shift
operator  while the  BEC is a trapped  in long  Rayleigh range  optical
dipole trap. The Hadamard rotation and a unitary shift operator on a BEC
prepared  in  one  of  the  internal  states followed  by  a  bit-flip
operation, implements  one step  of the Hadamard  walk.  To  realize a
sizable number of steps, the  process is iterated without resorting to
intermediate  measurement.   With current  dipole  trap technology,  it
should  be  possible  to  implement  enough  steps  to  experimentally
highlight  the discrete  quantum random  walk using  a BEC  leading to
further exploration of quantum random walks and its applications.
\end{abstract}

\maketitle
\preprint{Version}

\section{Introduction}

\label{intro}

Theoretical  studies  and the  experimental  possibility to  construct
quantum computers~\cite{nielsen-chuang}, based  on the laws of quantum
mechanics   has    been   around   for   about   two    and   a   half
decades~\cite{fey}. The Deutsch-Jozsa algorithm~\cite{DJ92}, communicated
in 1992, suggests that the  quantum computers may be capable of solving
some  computational problems  much more  efficiently than  a classical
computer. In  1994, Shor~\cite{shor} described  theoretically a quantum
algorithm to  factor large  semiprimes that was  exponentially faster
than the best-known  classical counterpart. Shor's algorithm initiated
active  research  in  the  area  of quantum  information  and  quantum
computation  across a  broad  range of  disciplines: quantum  physics,
computer sciences,  mathematics, and  engineering.  This research  has
unveiled  many new effects  that are  strikingly different  from their
classical counterparts. Since Shor's algorithm, Grover in 1997 devised
an algorithm~\cite{grover}, which can, in principle, search an unsorted
database quadratically faster than any classical algorithm.  Yet there
is  still a  search for  new quantum  search algorithms,  which  can be
practically  realized.  In this  direction,  adapting a known  classical
algorithm has  also been  considered as an option.   \par The
random    walk,    which    has    found    applications    in    many
fields~\cite{barber-ninham},  is  one  of  the aspects  of  information
theory  and  it  has  played   a  very  prominent  role  in  classical
computation.  Markov  chain  simulation  has  emerged  as  a  powerful
algorithmic tool~\cite{mark}  and many other  classical algorithms are
based on  random walks. It  is believed that exploring  quantum random
walks  allows,   in  a   similar  way,  a   search  for   new  quantum
algorithms. Quantum  random walks~\cite{kempe} have  been investigated
by   a   number  of   groups   and   several   algorithms  have   been
proposed~\cite{childs, shenvi,  childs1, ambainis}. To  implement such
an  algorithm in  a  physical system,  a  quantum walker  moving in  a
position and momentum space has  to be first realized. Some setups for
one-dimensional realizations  have  been   proposed  and  analyzed,
including trapped ions~\cite{travaglione} and neutral atoms in optical
lattice ~\cite{rauss}. One- and two-dimensional quantum walks in an array
of   optical  traps  for   neutral  atoms   has  also   been  analyzed
in Ref.~\cite{eckert}.  \par This paper proposes a scheme to experimentally
implement a discrete quantum  (Hadamard) walk using a coherent macroscopic
sample  of ultracold  atoms, a Bose-Einstein condensate  (BEC), in  one
dimension. The implementation  of the quantum walk using a BEC, gives access to
the quantum  phenomena on  a macroscopic scale.  With the  current BEC
trapping technique, it should be possible to implement enough steps to
experimentally  highlight  the quantum  walk  by generating  spatially
large-scale  entanglement and lead to the further exploration of the quantum
walk and its applications.  To implement the quantum walk, the BEC trapped
in  a  far  detuned  optical  dipole trap  with a long  Rayleigh  range
\footnote{Distance at which the diameter of the spot size increases by
a factor of $\sqrt 2$, $Z_{R}=\pi \omega_{0}^2/\lambda$ is the Rayleigh
range at wavelength $\lambda$ and beam waist $\omega_{0}$.} is made to
evolve  into the quantum superposition (Schr\"odinger cat) state of the two
trappable  states by applying the Hadamard  operation. The BEC in the
Schr\"odinger cat state is then  subjected to a unitary shift operator
to  translate the  condensate in  the axial  direction of  the optical
trap.  Thereafter, a compensatory bit-flip operator is applied.

This paper uses  rf pulses to implement the Hadamard  rotation and the
bit-flip operation, stimulated Raman transition technique as a
unitary shift operator to impart a well-defined momentum to
translate the BEC.  This method of applying a unitary shift
operator by subjecting it to optical Raman pulses to drive
transitions between two internal trappable states giving the BEC a
well defined momentum in an axial  direction of the trap can be
called {\it stimulated  Raman kicks}. The Hadamard operation
followed by a unitary shift operation and bit-flip operation
implements one step  of the quantum (Hadamard) walk.  A sizable
number of steps can be implemented  by iterating the process
without  resorting to the intermediate measurement.
\par
The  paper is  organized  as  follows.  Section \ref{qrw}  has a
brief introduction to the quantum random walk.  Section
\ref{bec-intro} discusses the quantum  walk using the BEC. The
creation of a Schr\"odinger cat state of the BEC, stimulated Raman
kicks used to  implement a unitary operator and the compensatory
bit flip  is discussed in  Secs. \ref{catstate},  \ref{ramankick},
and \ref{bitflip} respectively. Section \ref{phy-setup} discusses
the physical   setup required  for the implementation of quantum
walks  using the BEC. Section \ref{decohere} discusses decoherence
and  physical limitations and Sec. \ref{summary} concludes with
the summary.

\section{Quantum random walk}
\label{qrw}

Quantum random walks are the counterpart of classical random walks for
particles,  which   cannot  be  precisely  localized   due  to  quantum
uncertainties. The  word quantum random  walk, for the first time, was
coined in 1993 by Aharonov, Davidovich, and Zagury~\cite{aharonov}. In
the one-dimensional classical random walk, making a  step of a given
length to the left
or right for  a particle is described in  terms of probabilities.  On
the other hand, in the quantum random  walk it is described in terms of
probability amplitudes. In an  unbiased one-dimensional classical random
walk, with the particle initially at $x_{0}$, it evolves  in such a way
that at each step, the particle moves with probability 1/2 one step to
the left and with probability  1/2 one step to  the right. In  a quantum-mechanical analog, the state of the particle  evolves at each step into a
coherent superposition of moving one step to the right and one step to the left
with an equal probability amplitude.   \par Unlike the classical random walk,
two degrees of freedom are required for quantum random walks, internal
state  (superposition), which is  called {\it coin}  Hilbert space
$\mathcal H_{c}$  (quantum coin), and the {\it  position} Hilbert space
$\mathcal H_{p}$. Imagine a coin Hilbert space $\mathcal H_{c}$
of a  particle on a line  spanned by two basis  states $|0\rangle$ and
$|1\rangle$  and the position Hilbert  space $\mathcal  H_{p}$
spanned by  a basis state  $|x\rangle$:$x$={\bf Z}. The state  of the
total  system is in the  space $\mathcal  H= \mathcal  H_{c} \otimes
\mathcal  H_{p}$. The internal  state of  the particle  determines the
direction of the particle movement when the unitary shift operator $U$
is  applied on the  particle, \be  U=|0\rangle \langle  0|\otimes \sum
|x-1\rangle \langle x |+|1\rangle  \langle 1 |\otimes \sum |x+1\rangle
\langle x|.   \ee If the  internal state is $|0\rangle$,  the particle
moves to the  left, while it moves to the right  if the internal state
is                $|1\rangle$,     i.e.,
\bc
$U|0\rangle\otimes|x\rangle=|0\rangle\otimes|x-1\rangle$  \\
\ec
and \\
 \be
U|1\rangle\otimes|x\rangle=|1\rangle\otimes|x+1\rangle  \ee  Each step
of the quantum   (Hadamard)  walk  is  composed  of  the Hadamard  operation
(rotation) $H$,
\begin{equation}
H =\frac{1}{\sqrt 2} \left( \begin{array}{clcr}
 1  & &   1   \\
 1  & &  -1
 \end{array} \right),
\end{equation}

\noindent
on  the particle  to  bring  them to a superposition  state with  equal
probability,       such      that,
\bc
\[
(H \otimes \mathbbm{1})|0,x\rangle=\frac{1}{\sqrt 2}[|0,x\rangle+|1, x\rangle]
\] \\
\ec
and
 \be
(H \otimes \mathbbm{1})|1,  x\rangle=\frac{1}{\sqrt 2}[|0, x\rangle-|1,
x\rangle]
\ee
followed by a  unitary shift operation, $U$, which  moves the particle
in  the  superposition of  the  position  space.  If the  particle  is
prepared in the superposition state then each step of the quantum walk
consists of a unitary shift operator followed by  a Hadamard operation.
\par The probability amplitude  distribution arising from the iterated
application of $W=U(H \otimes \mathbbm{1})$ is significantly different
from  the distribution  of  the  classical walk  after  the first  two
steps~\cite{kempe}.   If   the  coin   initially  is  in   a  suitable
superposition  of  $|0\rangle$ and  $|1\rangle$,  then the  probability
amplitude distribution after  $n$ steps of the quantum walk  will have two
maxima symmetrically  displaced from the starting  point. The variance
of the quantum version  grows  quadratically with the number  of steps  $n$,
$\sigma^{2}\propto  n^{2}$ compared to  $\sigma^{2}\propto n$  for the
classical walk.
\section{Quantum Walk using the Bose-Einstein Condensate}
\label{bec-intro}

As discussed in sec. \ref{qrw},  two degree of freedom, the coin
Hilbert space  $\mathcal H_{c}$ and  the position  Hilbert space
$\mathcal H_{p}$ are required to implement the quantum walk.  A state of the BEC
formed from the atoms in one of the hyperfine states  $|0\rangle$  or $|1\rangle$
can be represented as $|0_{BEC}\rangle$ or $|1_{BEC}\rangle$ (the state of $N$ condensed atoms). The BEC formed is then transferred  to an optical  dipole   trap  with  long  Rayleigh   range $Z_{R}$. With the appropriate choice     of      power     and      beam     waist, $\omega_{0}$\footnote{Minimum radius of the beam, at the focal point.} of the trapping  beam, the condensate can remain  trapped at any point
within the  distance  $\pm Z$  from  the focal  point  in the  axial
direction of the  beam~\cite{chandra}. The BEC in the  optical trap is
then subjected  to evolve  into the superposition (Schr\"odinger cat)
state $\frac{1}{\sqrt 2}[| 0_{BEC}\rangle+i|1_{BEC}\rangle]$ (sec. \ref{catstate})
by applying the Hadamard rotation.  The coin Hilbert space $\mathcal
H_{c}$ can then be defined to be spanned by the two internal trappable
states   $|0_{BEC}\rangle$  and  $|1_{BEC}\rangle$   of  the   BEC.  The   BEC  in
the superposition state is spatially  translated by applying a unitary shift
operator, then  the position  Hilbert space $\mathcal  H_{p}$ is
spanned by the position of the  BEC in the long Rayleigh range optical
trap and is augmented by the coin Hilbert space $\mathcal H_{c}$.

The position of  the BEC trapped in the center of  the optical trap is
described by  a wave packet $|\Psi_{x_{0}}\rangle$  localized around a
position $x_{0}$, i.e., the  function $\langle x | \Psi_{x_{0}}\rangle$
corresponds to a wave packet centered  around $x_{0}$. When the BEC is
subjected     to     evolve      into     the     superposition     ($|S\rangle=[a|0_{BEC}\rangle+b|1_{BEC}\rangle]$) of the eigenstates $|0_{BEC}\rangle$ and
$|1_{BEC}\rangle$, its wave function is written as
 \be
\label{wavefunction}
|\Psi_{in}\rangle=\frac{1}{\sqrt 2}[|0_{BEC}\rangle+|1_{BEC}\rangle]\otimes|\Psi_{x_{0}}\rangle.
\ee
\noindent
where $|a|^2 + |b|^2=1$ and for symmetric superposition $a= b=
\frac{1}{\sqrt 2}$.
Once the BEC is in the superposition state, a unitary shift operator,
{\it stimulated Raman kick} (Sec. \ref{ramankick}), corresponding  to
one step length $l$ is applied;
\be
U^{\prime} = (\hat{X} \otimes \mathbb{I})\exp(-2iS_{z}\otimes Pl),
\ee
\noindent
$P$ being the momentum operator and $S_{z}$ the operator corresponding
to the step of length $l$. Step length $l$ is chosen to be less
than the spatial width of the BEC wave packet $|\Psi_{x_{0}}\rangle$.
\par
The application of the unitary shift operator ($U$) on the wave function of
Eq. (\ref{wavefunction}) spatially entangles the position and the
coin space and implements the quantum walk
\be
U^{\prime}|\Psi_{in}\rangle=\frac{1}{\sqrt 2}[|1_{BEC}\rangle\otimes e^{-iPl}
+|0_{BEC}\rangle\otimes e^{iPl}]|\Psi_{x_{0}}\rangle,
\ee
\noindent
where the wave packet is  centered around $x_{0}\pm l$.  Note that the
values  in the  coin  space have  been  flipped. This  is corrected  by
applying a compensating  bit flip on the BEC.   After each compensated
unitary    shift     operator,    $U    \equiv     (\hat{X}    \otimes
\mathbb{I})U^{\prime}$, the  BEC settles down  in one of  the internal
states  $|0_{BEC}\rangle$ or  $|1_{BEC}\rangle$. Therefore, a Hadamard
rotation is applied  to make the BEC  evolve into  superposition  state.
A  Hadamard rotation followed by a unitary  shift operator implements one
step of the Hadamard walk.   To realize the large  number of steps  the process is
iterated  without resorting  to intermediate  measurement in  the long
Rayleigh-range optical dipole trap.  \par

\subsection{Macroscopic cat state in the Bose-Einstein condensate}
\label{catstate}
Various   schemes  have  been   proposed  for   producing  macroscopic
superposition      or       Schr\"odinger      cat      states      in
BECs~\cite{cirac,ruostekoski, gordon}.  The scheme described  by Cirac
{\it  et al.}  in Ref.~\cite{cirac},  shows  that if  the  two species  are
Josephson coupled, then in  certain parameter regimes the ground state
of  the Hamiltonian  is  a  superposition of  two  states involving  a
particle  number  imbalance between  the  two  species.  Such a  state
represents a superposition of two states which are macroscopically (or
mesoscopically)   distinguishable,   and  hence   can   be  called   a
Schr\"odinger cat  state. Ruostekoski {\it  et al.}~\cite{ruostekoski}
have  also  shown that  such  states can  be  created  by a  mechanism
involving the  coherent scattering of far-detuned light  fields and it
neglects  the collisional  interactions between  particles.  The major
drawback of the  above schemes is that the time needed  to evolve to a
cat state  can be  rather long, and  thus problems due  to decoherence
would   be   greatly   increased.  In Ref.~\cite{cirac}   the   macroscopic
superposition  is produced  by  the normal  dynamic  evolution of  the
system. Gordon  {\it et al.}~\cite{gordon}  have proposed a  scheme in
which the superposition is produced by an adiabatic transfer of the ground
state  of  the Josephson-coupling  Hamiltonian,  that  is, after  the
initial state preparation the Josephson coupling is turned on for some
amount of time and turned off. The resulting modified quantum state is
a  Schr\"odinger  cat state.   But,  the  production  of such  a  state
using scheme~\cite{gordon} involves   considerable   experimental
difficulty.  \par The  ideal scheme to demonstrate the quantum  walk is to
confine the BEC that has an attractive interaction between  atoms in two
hyperfine  levels  $|0\rangle$ and  $|1\rangle$  in  a single  optical
potential well. The Hadamard operation (rotation) is applied on the BEC in
the  potential well  to transfer  (or rotate)  the atoms  part  of way
between  states $|0\rangle$  and   $|1\rangle$  using  a  resonant  rf
pulse \footnote{Microwave pulses are also used.} of duration $\tau$ and
detuning $\Delta$ from the  rf resonance~\cite{band}. The resonance rf
pulse couples the atomic hyperfine states $|0\rangle$ and $|1\rangle$ with
a coupling  matrix element $\hbar\omega_R/2$, where  $\omega_R$ is the
Rabi frequency and  the duration of pulse is much shorter than the
self-dynamics  of the  BEC.   The amplitude  of  these states  evolves
according to the Schr\"odinger equation,
\begin{equation}
\label{rf}
i \hbar \frac{d}{d\tau}\left(\begin{array}{clr}
a\\
b
\end{array}\right)=\hbar  \left( \begin{array}{clcr}
 0  & &  \omega_{R}/2   \\
 \omega_{R}/2  & & \Delta
 \end{array} \right)\left(\begin{array}{clr}
a\\
b
\end{array}\right).
\end{equation}

\noindent
At  this  stage,  each  atom  evolves  into  the  superposition  state
$a(\tau)|0\rangle+b(\tau)|1\rangle$,                               with
$a(\tau)=\cos(\omega_R\tau/2)$  and $b(\tau)=\sin(\omega_R\tau/2)$ for
detuning $\Delta=0$. The $N$-particle wave function of the BEC is a  product
of the single-particle superpositions of  $|0\rangle$ and  $|1\rangle$, that is,
it is still a microscopic superposition and is given by
\[
[a(\tau)|0\rangle+b(\tau)|1\rangle]^{N}=
\]
\be
\label{super}
\sum_{n=0}^N  \sqrt \frac{N!}{n!(N-n)!} a(\tau)^{N-n}b(\tau)^{n}|N-n, n\rangle,
\ee
where $|N-n, n\rangle$ is the state with $N-n$ atoms in state $|0\rangle$ and $n$ atoms in state $|1\rangle$. The individual atoms in the superposition state interact among themselves. Interatomic interactions, which provide nonlinear terms through binary collision as seen from the viewpoint of single-particle dynamics helps in generating highly entangled many body states~\cite{mandel, dalfovo}. The attractive inter-atomic interaction between  atoms in the superposition state induce entanglement between atoms. Thus Eq. (\ref{super}) evolves into the macroscopic superposition state in which  all  atoms  are   simultaneously  in
level  $|0\rangle$ and level  $|1\rangle$~\cite{cirac, dalvit1, mewes},
\be
[a(\tau)|0\rangle+b(\tau)|1\rangle]^{N}=
 =a(\tau)^{N}|N, 0\rangle + b(\tau)^{N}|0,N\rangle
\ee
where $|N, 0\rangle=|0_{BEC}\rangle$  and $|0, N\rangle=|1_{BEC}\rangle$.
Symmetric probability distribution $a(\tau)=b(\tau)=\frac{1}{\sqrt  2}$  can  be  obtained  by  carefully choosing $\omega_R$ and $\tau$.
\subsection{Unitary operator - Stimulated Raman kicks}
\label{ramankick}
A  unitary shift (controlled-shift) operation $U$ is applied on
the BEC  to spatially entangle the position and the coin space and implement
the quantum walk. Various schemes have been
worked out to give momentum kick to ultracold atoms in a
trap~\cite{mewes, phillips}. Mewes {\it et al.}~\cite{mewes} had
developed a technique where coherent rf-induced transitions were
used to change the internal state of the atoms in the magnetic
trap from a trapped to an untrapped state and thus displacing
the untrapped atoms from the trap. This method, however, did not
allow the direction of the output-coupled atoms to be chosen.
Hagley {\it et al.}~\cite{phillips} developed a technique to
extract sodium atoms from the trapped BEC by using a stimulated
Raman process between magnetic sublevels to have a control on the
direction of the fraction of the outgoing BEC and was used to
extract atom lasers from the BEC.

\begin{figure}
\begin{center}
\epsfig{figure=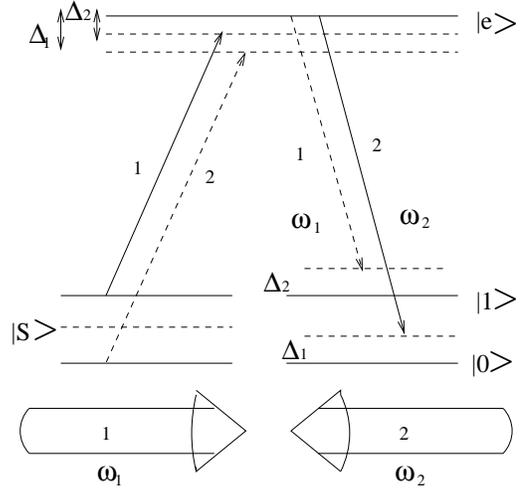, height= 6.5cm}
\caption{\label{lightfield}Light  field  configuration for a stimulated
Raman transition  process to give directional  momentum kick. $\Delta$
is the detuning of the laser from its transition frequency.
$|S\rangle$ signifies $(1/\sqrt{2})(|0\rangle \pm |1\rangle)$.}
\end{center}
\end{figure}

A  stimulated Raman  process  can  also be  used  to drive the
transition between  two optically  trappable states  of the atom $|0\rangle$
and $|1\rangle$ using the virtual state $|e\rangle$ as an intermediary
state, and impart a well-defined momentum to spatially  translate
the atoms in the coherent state (BEC). A unitary shift operation
$U^{\prime}$  thus  can  be applied  on  the atom in the BEC  using
a stimulated Raman process. A pair of counterpropagating laser beams
1 and   2  (Fig.~\ref{lightfield}) with  frequency   $\omega_{1}$  and
$\omega_{2}$ and wave vectors $k_{1}$ and $k_{2}$ is  applied on
the BEC for a $2N$-photon transition time ($N$ is the number  of atoms in the BEC)
to  implement one  unitary  shift  operation.   These beams  are
configured  to propagate  along  the axial  direction  of the
optical dipole trap. A stimulated Raman transition occurs when an
atom changes its  state by  coherently  exchanging photons
between  the two  laser fields,  absorption  of the photon  from
laser field  1  and  stimulated emission  into  laser  field 2  or
by  absorption  from field  2  and stimulated emission  into field
1.   The BEC initially in  eigenstates of the atom $|0\rangle$  ($|1\rangle$)
can  absorb  the photon from  field  1 (2)  and reemit the photon into
field 2  (1).  This  inelastic  stimulated Raman scattering
process imparts well-defined momentum on the coherent atoms,
\be
{\bf P}=\hbar (k_{1}-k_{2})=\hbar \delta z \ee
to the left during $| 0\rangle \xrightarrow {\omega_{1}}|e\rangle\xrightarrow{\omega_{2}}| 1 \rangle $ and
\be
{\bf P}= \hbar
(k_{2}-k_{1})=-\hbar \delta z
\ee
 to the right during $| 1\rangle \xrightarrow{\omega_2} |e\rangle \xrightarrow{\omega_1}  | 0 \rangle $, where
 \bc
  $| k_{1}-k_{2}| = |k_{2}-k_{1}|=\delta$,
\ec
\noindent
Thus, conditioned to being  in coin state  $|0\rangle$ ($|1\rangle$),
the  atoms in the BEC receive  a  momentum kick (shift)  $\hbar  \delta z$  ($-\hbar
\delta  z$).   This  process  of  imparting momentum  is  called
stimulated Raman kick and can  be analyzed as a photon absorption and
stimulated emission between  three bare states $|0\rangle$, $|1\rangle$,
and $|e\rangle$ driven by  two monochromatic light fields.  The matrix
element  for photon  absorption  and stimulated  emission during  each
stimulated  Raman kick  can  be written  as  \be \langle  n_{k_{1}}-1,
n_{k_{2}}+1, 1|\hat {H}_{I_{a}}|n_{k_{1}},n_{k_{2}}, 0 \rangle \ee
\noindent
during $| 0\rangle \xrightarrow {\omega_{1}}|e\rangle\xrightarrow{\omega_{2}}| 1 \rangle $ and
\be
\langle n_{k_{1}}+1, n_{k_{2}}-1, 0|\hat
{H}_{I_{b}}|n_{k_{1}},n_{k_{2}}, 1 \rangle
\ee
\noindent
during $| 1\rangle \xrightarrow{\omega_2} |e\rangle \xrightarrow{\omega_1}  | 0 \rangle $. ${H}_{I_{a}}$ and ${H}_{I_{b}}$ are the interaction Hamiltonians (electric dipole Hamiltonians). The field causing the transitions between bare states $|0\rangle$ and $|e\rangle$ is detuned from resonance by $\Delta_{1}$ and has a constant dipole matrix element  $V_{1}$ and phase $\varphi_{1}$. The field causing transitions between bare states $|e\rangle$ and $|1\rangle$ is detuned from resonance by $\Delta_{2}$ and has a constant dipole matrix element  $V_{2}$ and phase $\varphi_{2}$. The time-independent Hamiltonian in the rotating-wave approximation for this system can be written in terms of projection operations as
\[
\hat{H}_{I_{a}}=\hbar \Delta_1|e\rangle \langle e| + \hbar (\Delta_{1}+\Delta_{2})|1\rangle \langle 1|- \frac{\hbar V_{1}}{2} [|e\rangle\langle 0| exp(-i\varphi_1)
\]
\be
+|0\rangle\langle e| exp(i\varphi_1)]- \frac{\hbar V_{2}}{2} [|e\rangle\langle 1| exp(i\varphi_2)+|1\rangle\langle e| exp(-i\varphi_1)]
\ee
\noindent
In the above interaction picture, the energy of the bare  state $|0\rangle$ is chosen to
be zero.
\[
\hat{H}_{I_{b}}=\hbar \Delta_2|e\rangle \langle e| + \hbar (\Delta_{1}+\Delta_{2})|0\rangle \langle 0|- \frac{\hbar V_{2}}{2} [|e\rangle\langle 1| exp(-i\varphi_2)
\]
\be  +|1\rangle\langle   e|  exp(i\varphi_2)]-  \frac{\hbar  V_{1}}{2}
[|e\rangle\langle      0|      exp(i\varphi_1)+|0\rangle\langle     e|
exp(i\varphi_1)]. \ee  In the above  interaction picture the  energy of
the    bare    state  of the atom  $|1\rangle$    is    chosen    to   be    zero.
$V_{i}=\frac{\mu_{i}\cdot E_{0_{i}}}{\hbar}$, for $i={1,2}$. $\mu_{i}$
is  the  dipole  operator   associated  with  the  states  $|0\rangle$,$|e\rangle$
and $|e\rangle$,$|1\rangle$ and $E_{0}$  is the electric
field of the laser beam.

With the appropriate choice of  $V$, $\Delta$, and  $\varphi$, the probability  of
being  in bare  state  $|1\rangle$ ($|0\rangle$  depending on the  starting state)
after time  $t$ can be  maximized to be close to one, where $t$ is the time required for one  stimulated Raman kick. As discussed in Sec.~\ref{catstate}, after every stimulated Raman kick on atoms due to the attractive interaction between atoms, they evolve into the BEC state.

\subsection{Bit-flip operation}
 \label{bitflip}

The coin state of the atoms in the BEC  $|0\rangle$ ($|1\rangle$) after  stimulated
Raman kick  flips  to $|1\rangle$  ($|0\rangle$).  This  is  reversed
via a compensatory bit flip implemented using a resonant rf ($\pi$) pulse of
duration  $\tau_{\rm bf}$  and detuning  $\Delta$ from  the resonance.
The  amplitude  of  these  states  evolves, as  before,  according  to
Eq. (\ref{rf}).

\section{Physical setup for the implementation of the quantum walk using the Bose-Einstein Condensate}
\label{phy-setup}
A magnetic trap technique~\cite{bec} played a major role in the
first formation and early experiments on the BEC. But the magnetic
trap has a limitation to trap and manipulate atoms only of certain
sublevels. For $^{87}Rb$ atoms, the $|F=2,{\it m_{f}}= 2 \rangle$
state can be confined in the magnetic trap whereas $|F=1,{\it
m_{f}}= 1 \rangle$ cannot be confined using the magnetic
trap~\cite{feshbach}. Under appropriate conditions, the dipole
trapping mechanism is independent of the particular sub-level of
the electronic groundstate. The internal ground state can thus be
fully exploited using an optical dipole trap technique and be widely
used for various experiments~\cite{grimm}. Today the BEC from bosonic
atoms has been very consistently formed and manipulated using
various configurations of magnetic and optical  traps, an all
optical dipole trap technique has also been
developed~\cite{barrett}.

A BEC formed in one  of their internal states (eigenstates) $|0\rangle$
or  $|1\rangle$,   using  any  of   the  techniques~\cite{bec,  barrett}
mentioned above, is  transferred to a far detuned,  long Rayleigh range
($Z_R$) optical  dipole trap.  A  sizable number of steps  of the Hadamard
walk  can  be implemented  within  the  axial  range $\pm  Z$  without
decoherence in a long Rayleigh-range trap. Ref. ~\cite{chandra} has
calculated numerical values of potential depth, power required to trap
$^{87}Rb$  atoms at  distance $Z=x_{n}$  from the  focal point  of the
trapping beam  (after compensating for gravity) using  light fields of
various frequency and  beam waists $\omega_{0}$. In the  same way one
can work out required power and beam waists to trap ultracold atoms at
a distance $Z$  from the focal  point for different  species of  atoms using
laser fields of different detuning from the resonance.

Once the BEC is transferred into the optical dipole trap, a resonant rf
pulse (Hadamard rotation) of duration $\tau$ and detuning $\Delta$ is
applied to make it evolve into the Schr\"odinger cat state. The Schr\"odinger
cat state for the $^{87}Rb$ BEC trapped in one of the states
$|0\rangle$=$|F=1,{\it m_{f}}= 1 \rangle$ or $|1\rangle$=$|F=2,{\it m_{f}}= 2 \rangle$,
can be realized by applying a rf pulse, fast laser pulse, a standard
Raman pulse or microwave techniques.

The BEC in the Schr\"odinger cat  state is then subjected to a pair of
counterpropagating beams to implement the stimulated Raman kick for
duration  $t$,  $2N$  photon   transition  time  ($N$  being the number  of
photons). After implementing the stimulated Raman  kick the BEC is left to
translate for a  duration $P/ml$, the time taken by the condensate to
move  distance  $l$ during  or  after which  the  rf  $\pi$ pulse  is
applied.  This  process of applying  a rf pulse  (Hadamard rotation) and a
stimulated Raman  kick (shift), followed by a compensatory  rf pulse (bit
flip) is  iterated throughout  the trapping range  to realize  a large
number of steps (Fig.~\ref physical-qrw).

\begin{figure}
\begin{center}
\epsfig{figure=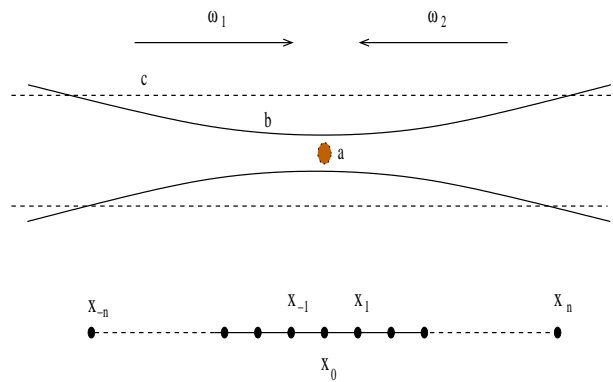, height= 5.0cm, width=8.0cm}
\caption{\label{physical-qrw}Physical setup  to implement the quantum walk
using the BEC. In the figure {\bf a} is a BEC, {\bf b} is a dipole trap with
a long Rayleigh range, and {\bf c} is the counterpropagating laser beam,
$\omega_1$  and $\omega_2$  used to  implement the stimulated  Raman kick,
unitary  operator.  A  Hadamard   rotation and Raman  pulse are  applied
throughout the trap region. The stimulated Raman pulse and the rf pulse are
applied alternatively to realize the Hadamard walk.}
\end{center}
\end{figure}
\begin{figure}
\begin{center}
\epsfig{figure=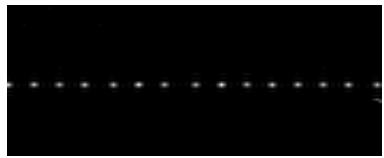, height= 2.0cm, width=5.0cm}
\caption{\label{dipole}Multiple microtraps to confine the BEC after it implements
a considerable number of steps of the quantum walk.}
\end{center}
\end{figure}
\subsection{Measuring the one-dimensional quantum walk probability distribution}
\label{prob}

After  $n$ steps  of the one-dimensional Hadamard walk  the superposition  in position
space is made  to collapse by applying multiple  microtraps to confine
and locate  the position  of the BEC.   The multiple microtraps  in a
line are created using the known technique ~\cite{dumke, ourpaper} (Fig.
~\ref{dipole}) and are switched  on,  removing the  long  Rayleigh  range
optical trap  simultaneously after  a time interval  of $T$.   \bc $T=
nt+(n\tau  -1)+n+nP/ml +  (n\tau_{\rm bf}-1)$.   \ec $T$  is  the time
taken by the BEC to travel $n$ quantum steps each of distance $l$ in a
line. The  time $t$  is the $2N$-photon  transition time  required to
implement one  unitary operation, $\tau$  is the time duration  of the
pulse used to bring the BEC to the superposition state, $P/ml$ is the time
taken by the BEC to move  one step distance $l$ with momentum $P$, and
$\tau_{\rm bf}$ is  the time duration required to  implement bit flip.
Note that  this time can  be eliminated if  the bit flip  is performed
during the translation of the BEC.

The spacing between each potential well is designed to be equal to the
distance  $l$.   The  condensate,  which has  undergone the Hadamard  walk,
is confined in  one of  the traps of  the multiple  microtrap when  it is
turned on.  Fluorescence measurement is performed on the condensate in
the microtrap to  identify the  final position of  the BEC  on which
the Hadamard walk has been implemented.  By repeating the experiment for a
fixed number of  steps (time $T$), the probability  distribution of the
position of the BEC after $n$ steps of the Hadamard walk can be obtained.

\section{Decoherence and physical limitations}
\label{decohere}
The decoherence of the BEC state also leads to the decoherence
of the quantum walk. When the atoms in the trap are not coherent they no longer
follow the coherent absorption and stimulated emission of light. Some
atoms absorb light field 1 and emit light field 2 and bring atoms to
state $|1\rangle$ and some absorb light field 2 and emit light field 1
and bring atoms to state $|0\rangle$, displacing atoms in both directions
in space giving no signature of displacement in the superposition of position
space. This will contribute to collision and heating,  and finally, atoms
escape out of the trap. The number of quantum steps that can be implemented
using the BEC without decoherence depends mainly on (a) the Rayleigh range,
as the BEC moves away from the trap center the width of the BEC wavepacket
increases and contributes to the internal heating of the atoms. Beyond a
certain distance $x_n$ from the dipole trap, center atoms in the trap
decohere resulting in the collapse of the quantum behavior. (b) The
stimulated Raman kick and rf pulse used to implement the Hadamard walk also
contribute to the internal heating and decoherence of the atoms after a few
iterations of the Hadamard walk. To implement a Hadamard walk, the spatial width of
the particle has to be larger than the step length $l$~\cite{aharonov}
and hence, $l$ for the Hadamard walk using the BEC is fixed depending on the spatial
width of the BEC used to implement the quantum random walk. However, decoherence
of the Hadamard walk does not happen even if some fraction atoms are lost from
the  BEC (and also untill the width of the wavepacket is larger than the step length).
\par
With the careful selection of beam waist and laser power one can have
a trap with a long Rayleigh range $Z_R$ ~\cite{chandra} so that the one-dimensional
quantum walk can be implemented in a line of close to one centimeter.

\section{Summary and conclusion}
\label{summary}
This paper proposes the use of the BEC in an optical dipole trap to
implement the quantum (Hadamard) walk in one dimension. The quantum
walk is performed in position space by periodically applying the
stimulated Raman kick and manipulating the internal state of the
BEC by using a rf pulse, Raman pulse, or  a microwave technique. After
$n$ steps the multiple microtrap is applied to confine the BEC
and fluorescence measurement is done to locate its final position.
To implement the Hadamard walk the width of the wavepacket has to be
larger than the step length $l$ and the width of the BEC wavepacket
is significantly larger than the other quantum particles suggested
to implement the Hadamard walk. This is one of the major advantages of
using the BEC to implement the Hadamard walk. With the presently available
efficient trapping technology, realizing the quantum random walk using
the BEC in a length of close to one centimeter (few 100 steps) is
expected. Realizing the quantum random walk using the BEC will be of
particular interest as the BEC is a macroscopic wavepacket. This
will give access to understanding the phenomena in a macroscopic
scale.
\\
\\
\bc
{\bf Acknowledgments :} \\
\ec
The author would like to thank Professor Keith Burnett for
his initial guidelines to start on this work and R. Srikanth, S.K. Srivatsa,
and Shrirang Deshingkar for helpful suggestions.


\end{document}